\documentclass[aps,prd,twocolumn,superscriptaddress]{revtex4-2}
\pdfoutput=1

\usepackage{graphicx,bm,amsmath,amssymb}
\usepackage{braket}

\usepackage{hyperref}
\usepackage{mathrsfs}
\usepackage{float}

\allowdisplaybreaks

\usepackage{tikz}
\newcommand{\drawsquare}{%
\tikz[scale=1,baseline=-\the\dimexpr\fontdimen22\textfont2\relax]{
\draw [fill=white] (-3pt,-3pt) rectangle (3pt,3pt);
}}
\newcommand{\drawblacksquare}{%
\tikz[scale=1,baseline=-\the\dimexpr\fontdimen22\textfont2\relax]{
\draw [fill=black] (-3pt,-3pt) rectangle (3pt,3pt);
}}
\newcommand{\drawcircle}{%
\tikz[scale=1,baseline=-\the\dimexpr\fontdimen22\textfont2\relax]{
\draw[fill=white] (0,0) circle (3pt);
}}
\newcommand{\drawblackcircle}{%
\tikz[scale=0.8,baseline=-\the\dimexpr\fontdimen22\textfont2\relax]{
\draw[fill=black] (0,0) circle (4pt);
}}
\newcommand{\drawtriangle}{%
\tikz[scale=1,baseline=-\the\dimexpr\fontdimen22\textfont2\relax]{
\draw[fill=white,yshift=-1pt] (90:4pt) -- (210:4pt) -- (330:4pt) -- cycle;
}}
\newcommand{\drawblacktriangle}{%
\tikz[scale=1,baseline=-\the\dimexpr\fontdimen22\textfont2\relax]{
\draw[fill=black,yshift=-1pt] (90:4pt) -- (210:4pt) -- (330:4pt) -- cycle;
}}

\usetikzlibrary{decorations.pathreplacing}
\newcommand{\drawparenthesis}{%
\tikz[scale=0.8,baseline=-\the\dimexpr\fontdimen22\textfont2\relax]{
\draw[align=center,decorate,decoration={brace,amplitude=5pt,raise=0pt},anchor=center] (0,-1) -- (0,1);
}}

\begin{document}

\title{Symmetry Constrained Quantum Error Mitigation for the Schwinger Model}

\author{Alexander Tomlinson}
\email{A.Tomlinson@soton.ac.uk}
\affiliation{Department of Physics and Astronomy, University of Southampton, Southampton, UK}

\author{Graham Van Goffrier}
\email{gwvg1e23@soton.ac.uk}
\affiliation{Department of Physics and Astronomy, University of Southampton, Southampton, UK}
\affiliation{Infleqtion, Unit 1, Oxford Technology Park, Technology Drive, Kidlington OX5 1GN, United Kingdom}

\author{Bipasha Chakraborty}
\email{B.Chakraborty@soton.ac.uk}
\affiliation{Department of Physics and Astronomy, University of Southampton, Southampton, UK}

\author{Zhenyu Cai}
\email{cai.zhenyu.physics@gmail.com}
\affiliation{Department of Computing, Imperial College London, 180 Queen’s Gate, London SW7 2AZ, United Kingdom}
\affiliation{Department of Engineering Science, University of Oxford, Parks Road, Oxford OX1 3PJ, United Kingdom}
\affiliation{Quantum Motion, 9 Sterling Way, London N7 9HJ, United Kingdom}

\begin{abstract}
Quantum error mitigation (QEM) is at the very heart of near-term quantum simulations and lattice gauge theories are no exceptions, rather their physical symmetries provide natural consistency checks on noise quantum states. In this work, we exploit the parity and fermion-number symmetries of a gauge theory,  the (1+1)-dimensional Schwinger model, under depolarising noise and investigate symmetry verification under digital quantum simulation. We investigate two set-ups - symmetry-sector post-selection in adiabatic state preparation followed by real-time measurements of the chiral condensate and symmetry verification within a variational quantum eigensolver (VQE). In the first case, post-selection reduces the bias in the chiral condensate consistently removing up to ~60\% of the quantum noise induced error in our system. Motivated by the observed regularity of the residual bias (in the low-noise regime), we further introduce a global-noise calibration obtained from classically accessible smaller lattices and implemented on larger lattices recovering noiseless chiral condensate values within statistical uncertainty. However, in VQE, symmetry verification does not seem to generally improve the optimised parameters or the fidelity of the prepared states, although it reduces the bias in the estimated ground-state energy. This demonstates that improving a noisy cost-function estimator in variational algorithms does not necessarily improve the outcome of the algorithm. Our results show the strength of symmetry verification in different approaches while establishing that its usefulness critically depends on where it is applied in the computational workflow, and provide practical guidance for symmetry-assisted quantum error mitigation in quantum simulations of lattice gauge theories. 
\end{abstract}
\maketitle
\section{Introduction}

Classical computing has been instrumental to the exploration of particle physics. One notable case is with lattice quantum chromodynamics, where discretising spacetime onto a lattice provides a non-pertubative method of calculation in strong coupling regimes where analytic solutions are intractable \cite{FLAG2024}. Reliant on Monte Carlo methods to sample the Euclidean path integral, lattice QCD is impeded by the infamous sign problem if the integrand is highly oscillatory \cite{Troyer2005}. Thus if one wants to investigate, for example, models with chemical potentials \cite{deForcrand2009}, the sign problem needs to be circumnavigated. One approach is to work directly with the Hamiltonian \cite{Kogut75}, however handling the exponentially growing Hilbert space as system size increases quickly runs into memory and runtime problems with classical computing. Quantum computing is hoped to offer advantage, although the current so-called NISQ-era devices are only at an order of ~100-1000 qubits and suffer from noisy gates \cite{Preskill2018}. Eventually it is hoped that full fault tolerance will be achieved, as the threshold theorem describes how quantum error correction becomes viable once error rates fall below a certain level \cite{Gottesman2009}. Hardware is constantly developing towards this goal, at which point potential circuit depths would jump significantly, however it is interesting to consider what may be achievable on the intermediate hardware. Quantum error mitigation is a nearer-term methodology which endeavours to reduce the effect of noise in this current era without full error correction \cite{Cai2022}. This allows one to either produce a stronger signal for the same circuit depth, or access deeper circuits with a target uncertainty level. It is naturally not possible to gain any improvement for free, so any method will have an associated overhead through either additional ancillary qubits, extra quantum operations, and/or the need for more samples. This should be carefully considered when evaluating efficacy. Many different techniques have been developed over the years. For example, zero-noise extrapolation takes advantage of taking readings with different error rates in order to extrapolate the trend to an ideal measurement \cite{PhysRevX.7.021050}. Probabilistic error cancellation seeks to explicitly use noisy operations, of which a linear combination can be taken to express the noiseless expectation value \cite{PhysRevLett.119.180509}. Some techniques target specific occurences of noise, such as dynamical decoupling which looks to suppress decoherence of idling qubits by inserting pulses which on average cancel out the interactions between the qubits and their environment \cite{PhysRevApplied.20.064027}. Measurement errors can be targeted using twirled readout error extinction which uses Pauli gates to convert arbitrary noise into a more predictable structure \cite{PhysRevA.105.032620} (Pauli twirling can also be used for in-circuit operations \cite{PhysRevA.88.012314}). In fact, popular software tools such as IBM's Qiskit and Amazon's Braket have inbuilt tools which implement many of these and can be easily toggled on and off \cite{Javadi-Abhari2024,braket}. It is also possible to combine multiple different mitigation methods providing they are compatible with each other. 

We investigate a particular mitigation technique called symmetry verification \cite{PhysRevA.98.062339}, which does not feature as an in-built tool. Symmetry verification takes advantage of an expected symmetry sector the targeted measurement falls in by validating that the incidence of noise has not moved the final quantum state outside this sector. Unlike several of the other techniques mentioned previously, symmetry verification can be agnostic of the exact structure of the noise profile which thus does not need to be learned. The chosen symmetry is typically verified with one of two approaches. One (which is used in this work) is a post-selection method where an explicit measurement of the symmetry is taken using additional quantum gates and any runs which measure in the wrong sector are discarded. This means ideally your symmetry and Hamiltonian commute such that they can be simultaneously diagonalised in a shared basis to measure in (this is not the case for part of this work, but a workaround is described). The other is a post processing method where no additional circuit elements are required, but the calculated operator is a product of the problem Hamiltonian with the symmetry operator. This has the downside of requiring a larger sampling overhead, but could be of interest for future work. Previous studies have for example tested symmetry verification for molecular physics with the $\text{H}_2$ molecule \cite{PhysRevA.100.010302}, and optimisation problems \cite{Kakkar2022}, but there is little work investigating in the context of quantum field theory problems. This is a natural choice to consider due to the multitude of symmetries one would expect to be able to exploit in a particle physics model. In our case we target the Schwinger model, i.e. $(1+1)$-dimensional quantum electrodynamics. Keeping the model relatively simple allows us to keep circuit sizes realistic for NISQ-era devices, whilst maintaining the potential to explore features of interest such as chiral symmetry breaking and confinement \cite{PhysRevD.13.1043}. Specifically, we opt to verify either parity or baryon number as these are easy to check with few gates, which is important not just to keep overhead to a minimum, but also since the verification process itself can incur noise which will not be mitigated. Regarding algorithms, we test symmetry verification with ground state preparation using two different approaches - variational quantum eigensolver (VQE) \cite{Peruzzo2014} and adiabatic state preparation (ASP) \cite{Albash2018}. VQE is a hybrid quantum-classical algorithm which measures the Hamiltonian expectation of a parametrised quantum circuit and then uses classical computation to optimise the parameters to a minimum in energy. These parameters can then be used in the circuit to produce an approximation of the ground state. ASP uses the adiabatic principle by starting the quantum system in the ground state of a known Hamiltonian, and evolves the system to the target Hamiltonian slowly enough such that the system remains in the ground state. This in theory, unlike VQE, can produce a ground state to arbitrary precision, at the cost of needing much longer coherent times. Therefore we can cover both a NISQ-era option in VQE and beyond NISQ-era with ASP. We take the ASP investigation one step further by time-evolving the prepared ground state in order to measure the chiral condensate. Our results show that for VQE whilst measurements of the Hamiltonian expectation do have a reduced bias, this does not significantly affect the optimisation process, and thus symmetry verification may not justify its required overhead. On the other hand, symmetry verification does demonstrate consistent improvement for the chiral condensate measurements (especially when a further fitting scheme is utilised), and thus we highlight that care should be taken when deciding to apply mitigation techniques.

\section{Quantum Simulation of Lattice Gauge Theories}\label{Sec:Schw}

We consider the problem of simulating the Schwinger model, which is a $1+1$-dimensional $U(1)$ gauge theory with the Lagrangian
\begin{equation}
\mathcal{L} = -\frac{1}{4}F_{\mu\nu}F^{\mu\nu}+i\overline{\psi}\gamma^{\mu}(\partial_{\mu}+igA_{\mu})\psi - m\overline{\psi}\psi,
\end{equation}
where $\psi$ is a Dirac fermion with mass $m$, the gauge coupling is given by $g$, and $F_{\mu\nu}=\partial_{\mu}A_{\nu}-\partial_{\nu}A_{\mu}$. We place this theory on a spatial lattice of $N$ sites with spacing $a$ using the staggered fermion formulation with an open boundary condition.
\begin{multline}
H = -iw \sum_{n=1}^{N-1} \left[\chi^{\dag}_n e^{i\phi_n}\chi_{n+1} - \textrm{h.c.}\right] \\
+ m \sum_{n=1}^N \left(-1\right)^n \chi^{\dag}_n \chi_n + J \sum_{n=1}^{N-1} L^2_n,
\end{multline}

where now $w = 1/(2a)$ and $J=g^2a/2$. The gauge operator $\phi_n \leftrightarrow -agA^1(x)$ lives on the $n$th site, whilst the link $L_n \leftrightarrow -\Pi (x)/g$, acts between sites $n$ and $n+1$.
The qubit mapping is achieved through the Jordan-Wigner transformation

\begin{equation}
\chi_n = \left(\prod_{l < n} -iZ_l\right) \frac{X_n -iY_n}{2},
\end{equation}
where $(X_n,Y_n,Z_n)$ corresponds to a Pauli gate on qubit $n$. Since we only have one spatial dimension, we can solve the Gauss law
\begin{equation}
L_n - L_{n-1} = \chi_n^{\dag}\chi_n - \frac{1-(-1)^n}{2},
\end{equation}
and redefine $\chi_n \rightarrow \prod_{l<n} [e^{-i\phi_l}] \chi_n$ to eliminate the gauge operators and express the model solely with spin operators in the form
\begin{equation}
H = H_{ZZ}+H_{\pm}+H_Z,
\end{equation}
where
\begin{align}
H_{ZZ} &= \frac{J}{2} \sum_{n=2}^{N-1} \sum_{1\leq k < l \leq n} Z_k Z_l \nonumber
\\
H_{\pm} &= \frac{w}{2} \sum_{n=1}^{N-1}[X_n X_{n+1} + Y_n Y_{n+1}] \nonumber
\\
H_Z &= \frac{m}{2}\sum_{n=1}^N (-1)^n Z_n - \frac{J}{2} \sum_{n=1}^{N-1} (n\bmod 2) \sum_{l=1}^n Z_l,
\end{align}
where constant terms have been omitted.

\section{Quantum Error Mitigation and Symmetry Verification}

The core objective for error mitigation is to reduce the bias introduced to measured expectation values from noisy hardware. This is achieved by post-processing the combined output of multiple circuit runs, in contrast to quantum error correction which is able to handle any individual run. For a desired readout accuracy, circuit depth must be balanced with the given hardware noise level such that the signal is is not lost. Mitigation allows for deeper circuits to be run in the case of worse noise to maintain this accuracy. This highlights the advantage over error correction, as there is no threshold error rate the hardware must achieve for the method to be viable. Furthermore, mitigation can stay relevant as machines improve, accessing deeper and deeper circuits.
\\
There are naturally costs to consider, due to a hardware overhead modifying the circuit with ancillary qubits and gates, a sampling overhead increasing the number of shots required, or a combination of the two. 

\subsection{Modelling Noise}

As this paper focuses on simulation of quantum hardware, it is important to establish the noise model implemented. Throughout this paper we adopt the notation and conventions of \cite{Cai2022}. The circuit fault rate $\lambda$ gives the average number of faults per circuit run, effectively quantifying the total amount of noise. We make the assumption that a fault may occur at any gate with equal error rate $p$, such that with $M$ total gates we have $\lambda=Mp$. We associate an output state with the relevant circuit fault rate as $\rho_{\lambda}$, and in this sense we denote the ideal state as $\rho_0$. The effect of noise is modelled using quantum channels, which are linear, completely positive and trace preserving maps. We represent the occurrence of errors using Kraus operators $K$, which evolve the density matrix as
\begin{equation}
\rho \rightarrow \sum_i K_i \rho K_i^{\dagger},
\end{equation}
where each $K_i$ has some associated probability. We make the common choice of using depolarising noise, which is modelled by the Kraus operators
\begin{align}
K_0 &= \sqrt{1-p} 
	\begin{pmatrix}
	1 & 0 \\
	0 & 1
	\end{pmatrix} \nonumber
\\
K_1 &= \sqrt{\frac{p}{3}} 
	\begin{pmatrix}
	0 & 1 \\
	1 & 0
	\end{pmatrix} \nonumber
\\
K_2 &= \sqrt{\frac{p}{3}} 
	\begin{pmatrix}
	0 & -i \\
	i & 0
	\end{pmatrix} \nonumber
\\
K_3 &= \sqrt{\frac{p}{3}} 
	\begin{pmatrix}
	1 & 0 \\
	0 & -1
	\end{pmatrix},
\end{align}
where one can consider the effect to be a Pauli causing either a bit flip, phase flip, or combination of the two. This is a popular choice as depolarising noise transitions the channel to the maximally mixed state, so can be considered a worst-case scenario where all information is lost.
\\
On top of reducing bias in the operator expectation $\mathrm{Tr}[O\rho_\lambda]$, another useful metric for performance is the fidelity boost, which compares the overlap with the ideal state between mitigated and unmitigated outputs as
\begin{equation}
B_{fid} = \frac{\mathrm{Tr}[\rho_0 \rho_{em}]}{\mathrm{Tr}[\rho_0 \rho_\lambda]},
\end{equation}
where $\rho_{em}$ is the mitigated state.

\subsection{Symmetry Verification}

There are many types of quantum error mitigation that have been explored. Some examples include zero noise extrapolation \cite{PhysRevX.7.021050}, where the expectation value is calculated at different $\lambda$ and fitting a curve, probabilistic error cancellation \cite{PhysRevLett.119.180509} where an ideal channel is constructed using a linear combination from a basis of noisy operations, and measurement error mitigation \cite{PhysRevA.103.042605} where information about the transition matrix between a binary string $y$ to $x$ at the point of measurement is learnt and inverted. We opt to investigate symmetry verification \cite{PhysRevA.98.062339}, where the occurrence of errors are detected by checking the output state against an expected symmetry and discarding any which fall outside the symmetry sector. This is an appealing choice to apply to gauge theory simulation, due to the inherent symmetries that can be exploited. The hardware overhead involves the additional circuit elements required to measure the symmetry, whilst the sampling overhead is the extra shots required to offset the discarded runs. Unlike, for example, probabilistic error cancellation, it is not possible to fully remove the noise profile, as faults which keep the system in the correct symmetry sector will not be detected, and it is also possible that the symmetry checking elements of the circuit incur noise themselves. That said, symmetry verification can be simple to implement, and the sampling overhead has reasonable scaling as the inverse of the fraction of passing circuit runs. It is also possible to implement as part of a package with other methods.
\par
The symmetry $S$ is chosen such that with single-qubit measurements to measure both the symmetry and target observable simultaneously, i.e. they share a basis in which they are diagonal. It should be noted that it is also possible as an alternative to fold the symmetry into the observable, $O_{sym}=\Pi O \Pi$ where $\Pi$ is the projector of the correct symmetry subspace, and solely measure this operator. Whilst simplifying the circuit, this comes at the cost of a higher sampling overhead, $C_{em} \sim \mathrm{Tr}[\Pi \rho]^{-2}$ compared with  $C_{em} \sim \mathrm{Tr}[\Pi \rho]^{-1}$. Since we will be using the easy to check parity operator symmetry, $S = \prod_i Z_i$ we opt to take advantage of the better sampling overhead of direct in-circuit verification.

\subsection{VQE}

A variational quantum eigensolver aims to take advantage of both quantum and classical computation in order to minimise the depth of the quantum circuit, to the extent NISQ-era devices can be targeted \cite{Peruzzo2014}. The quantum computation involves a parametrised ansatz circuit which generates a state of which the expectation of the target Hamiltonian can be measured. Since the ground state energy is guaranteed to be less than or equal to this measured energy, a classical computer can run an optimisation algorithm to find the parameters $\boldsymbol\theta$ which minimise the cost function,
\begin{equation}
C(\boldsymbol\theta) = \braket{\psi(\boldsymbol\theta) | H | \psi (\boldsymbol\theta)},
\end{equation}
which can then be reused to generate an approximation of the ground state. The Schwinger Model has been explored with VQE on up to 100 qubits \cite{PRXQuantum.5.020315}, and features such as the phase transition \cite{Angelides2025} and quench dynamics \cite{PhysRevD.108.034501} have also been investigated using the method. The goodness of approximation depends both on how well the parameter space captures the ground state, and how amenable the space is to navigate by the optimiser. If the Hamiltonian is composed of separate Pauli terms which do not commute, it is not possible to measure the full expectation in one run of state preparation since each qubit can only be measured in one basis before needing to be reset. Note this is on top of the need to run many shots to generate appropriate statistics. Any subset of terms which do commute can be determined in one run by measuring in their simultaneous eigenbasis. We use the qubit-wise commuting approach to find appropriate groups, where only operators acting on the same qubit need to commute. For the Schwinger model Hamiltonian, this trivially only gives three required measurements, where every qubit is measured in the Z-basis, then X-basis, then Y-basis.
\subsection{Ansatz}
We opt to use an ansatz designed by Gard et. al. which generates parametrised states respecting a chosen particle number \cite{Gard2020}. Thus, we can produce states which have half the total sites occupied, and implement a check that ensures after a noisy circuit the state is still in the correct parity sector. The ansatz is built up using iterations of a 2-qubit block of gates (referred to as an A gate), and is conjectured to span the targeted symmetry-respecting Hilbert subspace providing $\binom Nm$ A gates are used. The advantage of this ansatz is the ease in which it generalises to any number of qubits, however a significant roadblock is the number of parameters required as sites increase. This is due to the fact each A gate takes two parameters and thus the total number grows exponentially with system size. Imposing time reversal symmetry allows one parameter in each A gate to be fixed to 0, but still for our purposes (4, 6, 8) qubits requires (6, 20, 70) parameters. Optimisation is good in the first case, however the second and third cases suffer from the barren plateau problem, a typical issue with variational algorithms where the gradient of the cost function is prone to vanish. We opt to reduce the parameter space by using fewer A gates, and find a balance between how well the ansatz captures the ground state with how well the cost space can be optimised over. Through trial and error we found the most successful results were produced using 10 parameters for six qubits and 14 parameters for eight qubits.

\subsection{Mitigation Procedure}
\begin{figure}[h]
  \centering
  \includegraphics[width=.49\textwidth]{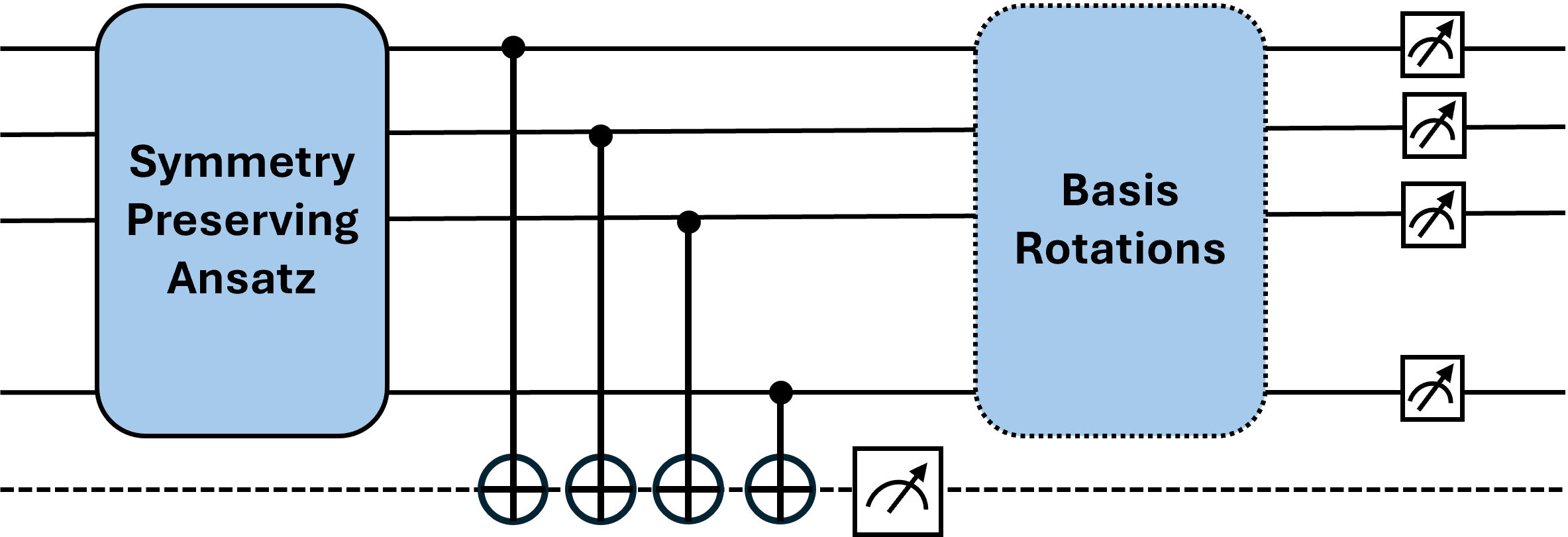}
  \caption{Structure of full circuit implementing error mitigation.}
  \label{fig:EM_VQE_Schematic}
\end{figure}

\begin{figure}[h]
  \centering
  \includegraphics[width=.49\textwidth]{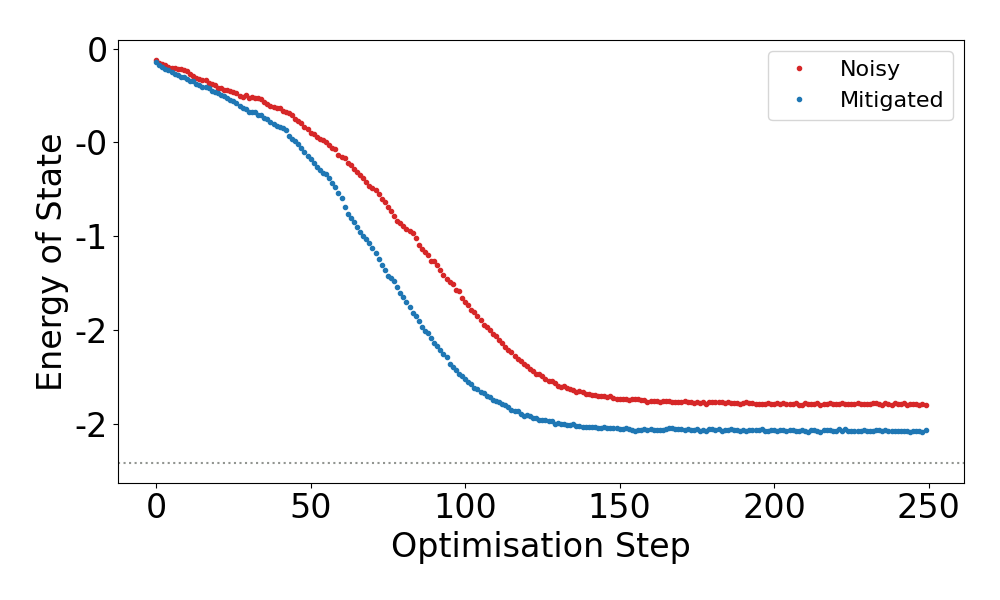}
  \caption{Comparison of noisy, and mitigated runs for $N=4$ qubits with $P(\text{error})=0.1.$}
  \label{fig:VQE_N4}
\end{figure}

We utilise a single ancilla to measure the symmetry, where the use of CNOT gates entangling every qubit with the ancilla allows us to determine the parity of the system, e.g. measuring the ancilla to be 0 indicates even parity. Unfortunately, as the full Hamiltonian is not diagonal in the computational basis, however as the measurements are divided into the qubit-wise commuting groups, the check can be made without issue for the $Z$ measurement run. For the $X$ and $Y$ runs we opt to implement the check before the measurement basis rotations. This means noise may be incurred after the check, however, as we only need at most two gates to rotate each qubit, this contribution will be small compared with the total noise brought about by the full circuit. This structure can be seen in Figure \ref{fig:EM_VQE_Schematic}. Through the use of this symmetry verification procedure at each VQE step, it is hoped that the improved accuracy in evaluating the cost function allows the optimiser to take better steps in the parameter space towards the ground state. Figure \ref{fig:VQE_N4} shows an example result for four qubits using an error rate of $0.1$, since for all three circuit sizes this gives roughly a unity circuit fault rate. Promisingly, it seems that mitigation improves the result, however this is misleading as shown in the following section. 

\subsection{Hybrid Results and Ideal State Overlap}
\begin{figure}[t]
  \centering
  \includegraphics[width=.49\textwidth]{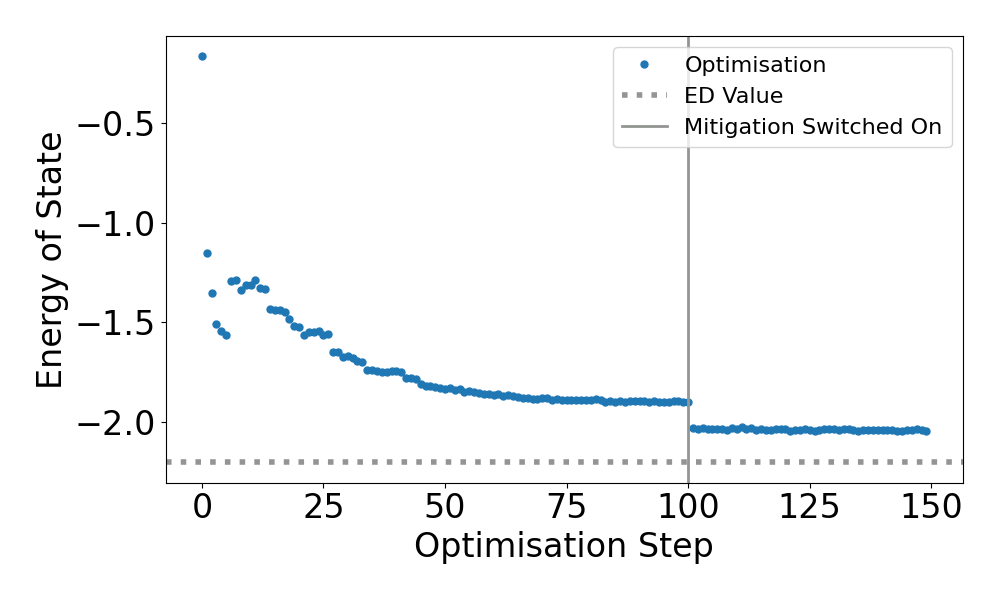}
  \caption{N=4 hybrid optimisation where error mitigation is only turned on after the first 100 steps}
  \label{fig:hybridVQE}
\end{figure}

\begin{figure}[t]
  \centering
  \includegraphics[width=.49\textwidth]{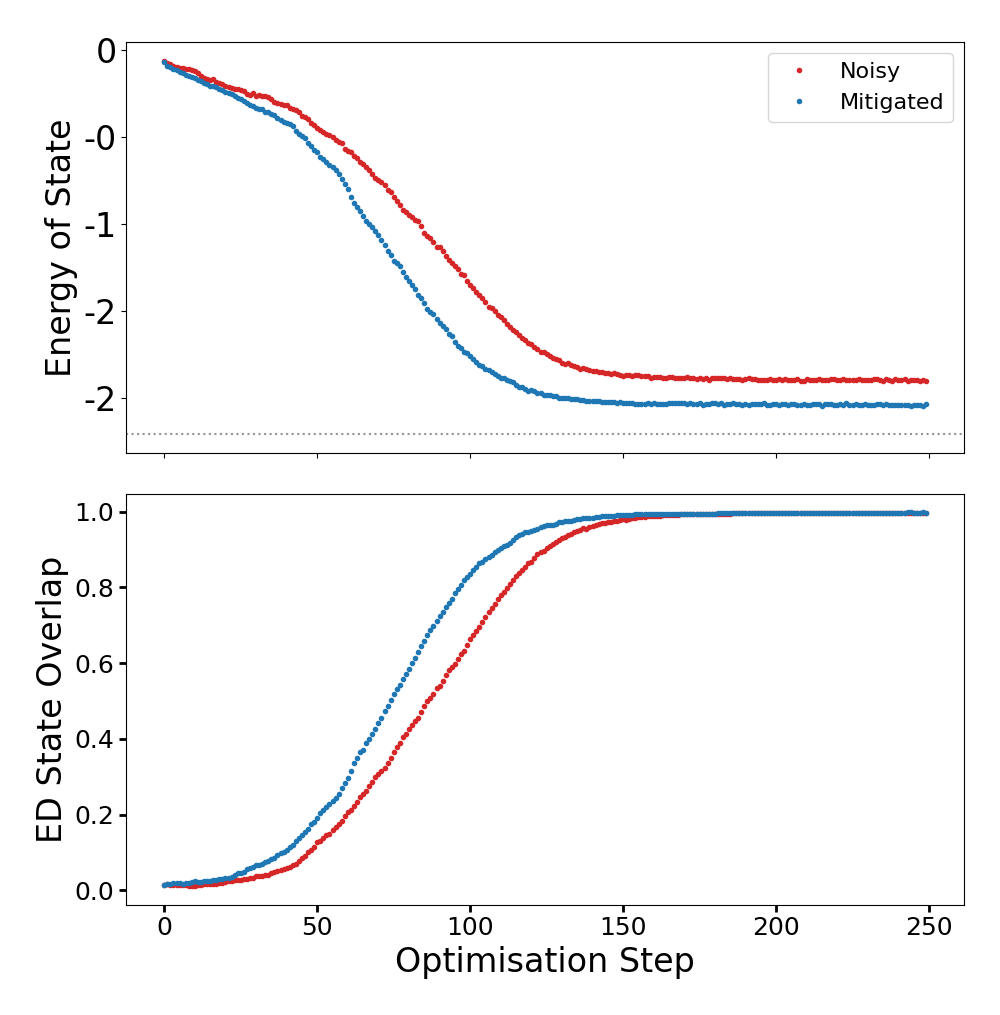}
  \caption{N=4 Gradient descent optimisation with same initial parameters showing comparison between states generated by mitigated or unmitigated parameters at each optimisation step}
  \label{fig:N4_Overlap}
\end{figure}

\begin{figure}[t]
  \centering
  \includegraphics[width=.49\textwidth]{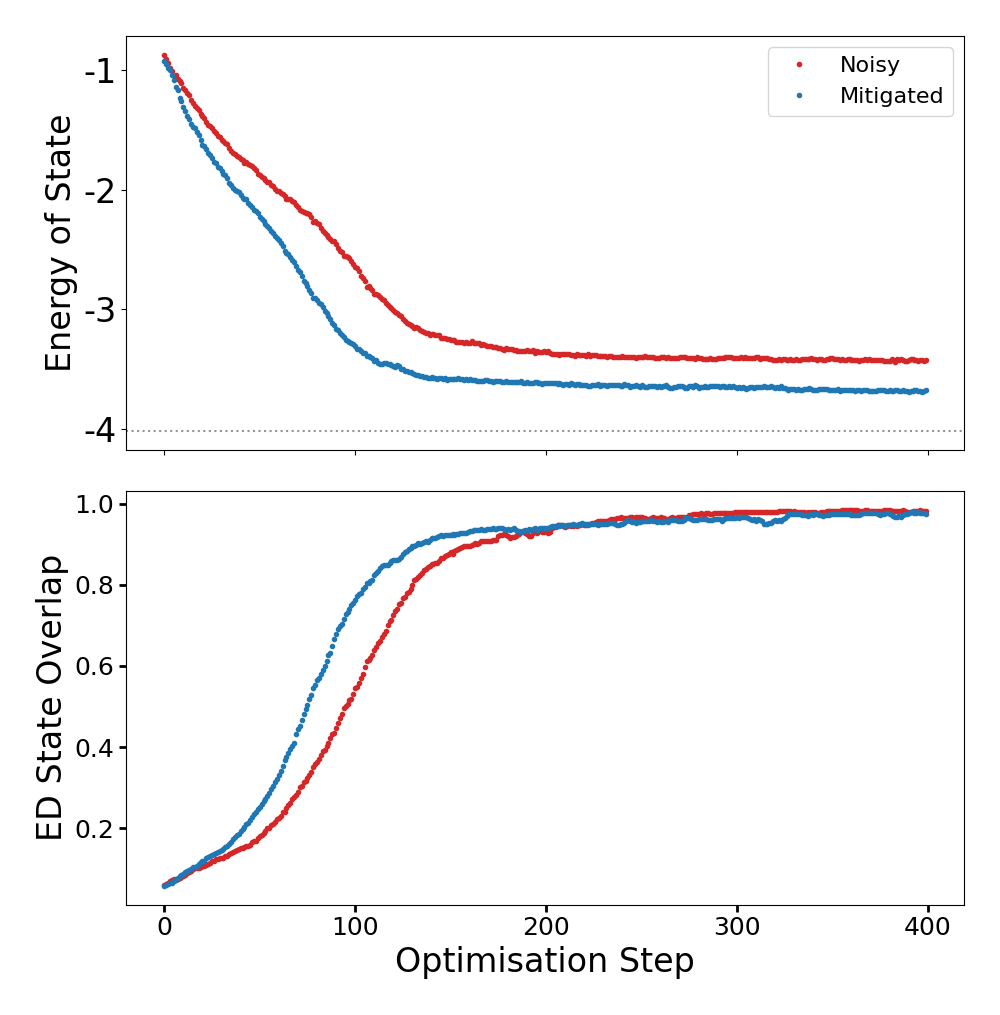}
  \caption{N=6 Gradient descent optimisation with same initial parameters showing comparison between states generated by mitigated or unmitigated parameters at each optimisation step}
  \label{fig:N6_Overlap}
\end{figure}

\begin{figure}[t]
  \centering
  \includegraphics[width=.49\textwidth]{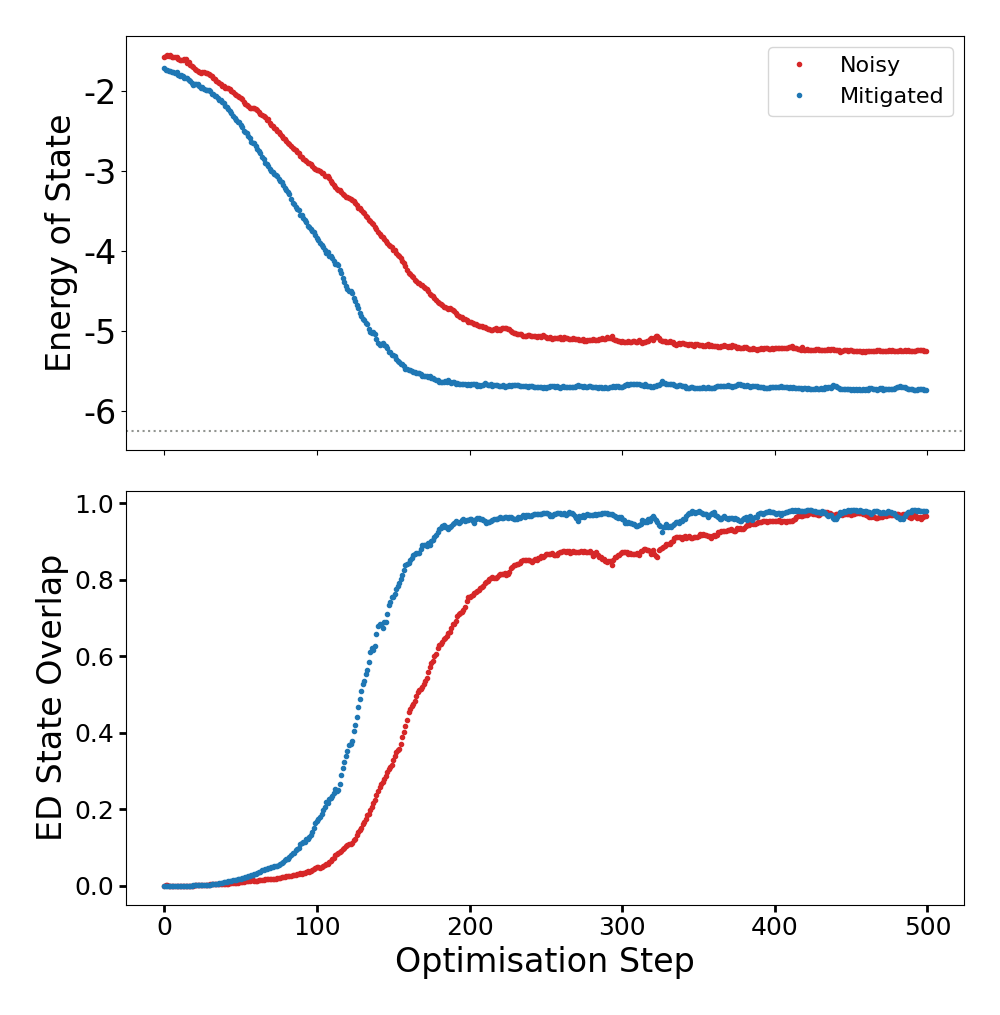}
  \caption{N=8 Gradient descent optimisation with same initial parameters showing comparison between states generated by mitigated or unmitigated parameters at each optimisation step}
  \label{fig:N8_Overlap}
\end{figure}

Implementing a hybrid approach gives stronger insight into the efficacy of parity symmetry verification during VQE optimisation. Specifically, the optimiser is run without mitigation such that it plateaus, at which point the mitigation technique is switched on and the same optimisation program continues. One might either expect no improvement (if worse optimisation has lead to a local minimum), or steady optimisation to a new plateau corresponding with the usual mitigated value. However, what is observed is an immediate discrete jump to the new value with no further optimisation, as can be seen in figure \ref{fig:hybridVQE}. This suggests that the parameters are not significantly updating further, rather a more accurate energy is calculated using the (roughly) same parameters. This is shown more explicitly in figures \ref{fig:N4_Overlap}, \ref{fig:N6_Overlap}, \ref{fig:N8_Overlap}, which compare separate mitigated and unmitigated runs by preparing the ansatz state from the parameters at each step in the optimisation by using a noiseless statevector simulator, and calculating the fidelity between the two states with an ideal state calculated using exact diagonalisation. As can be seen, it is consistently the case that overlap is high by the end of optimisation. This gives the hint to consider the fidelity boost, and table \ref{table:1} shows the average boost over the last fifty optimiser steps for 10 different runs. Concerningly, not only is the boost generally small, there are cases where performance is actually worse. In this case there is thus no practical gain implementing symmetry verification during VQE optimisation - it is only the final parameters which are of interest, the energy evaluation is only a useful metric to follow the optimisation process. This is not necessarily a surprise, as since the introduction of noise flattens the cost landscape, small-sized cases where the cost space is simple can still be navigated. One could be optimistic that as the cost space becomes more complicated (or noise levels increase), the benefit of symmetry verification might become more pronounced, but further investigation is required here.

\begin{table}[h!]
\centering
\begin{tabular}{|c|c|c|} 
 \hline
 N=4 & N=6 & N=8 \\ [0.5ex] 
 \hline\hline
 1.0007 & 0.9816 & 1.0000 \\ 
 \hline
 0.9997 & 1.0060 & 1.0022 \\
 \hline
 0.9996 & 1.0031 & 0.9808 \\
 \hline
 1.0014 & 1.0008 & 1.1718 \\
 \hline
 1.0013 & 1.0023 & 1.0193 \\
 \hline
  0.9995 & 1.0122 & 1.0064 \\
 \hline
  0.9978 & 1.0069 & 1.0510 \\
 \hline
  1.0016 & 0.9996 & 0.9200 \\
 \hline
  1.0045 & 1.0140 & 1.0642 \\
 \hline
 1.0013 & 1.0104 & 1.0339 \\ [1ex] 
 \hline
\end{tabular}
\caption{Fidelity boost for ten independent runs of $N=4, 6, 8$ parity verification VQE using gradient descent.}
\label{table:1}
\end{table}

\begin{figure}[h]
  \centering
  \includegraphics[width=.49\textwidth]{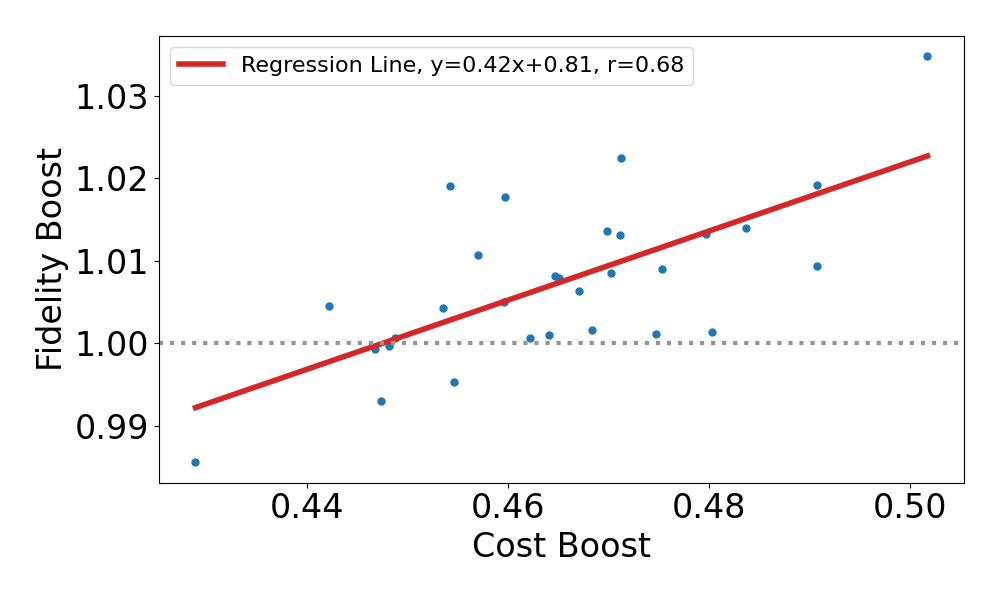}
  \caption{Correlation between the improvement of state fidelity from unmitigated to mitigated with the improvement to cost evaluation from unmitigated to mitigated for $N=4$}
  \label{fig:N4_Correlation}
\end{figure}

\begin{figure}[h]
  \centering
  \includegraphics[width=.49\textwidth]{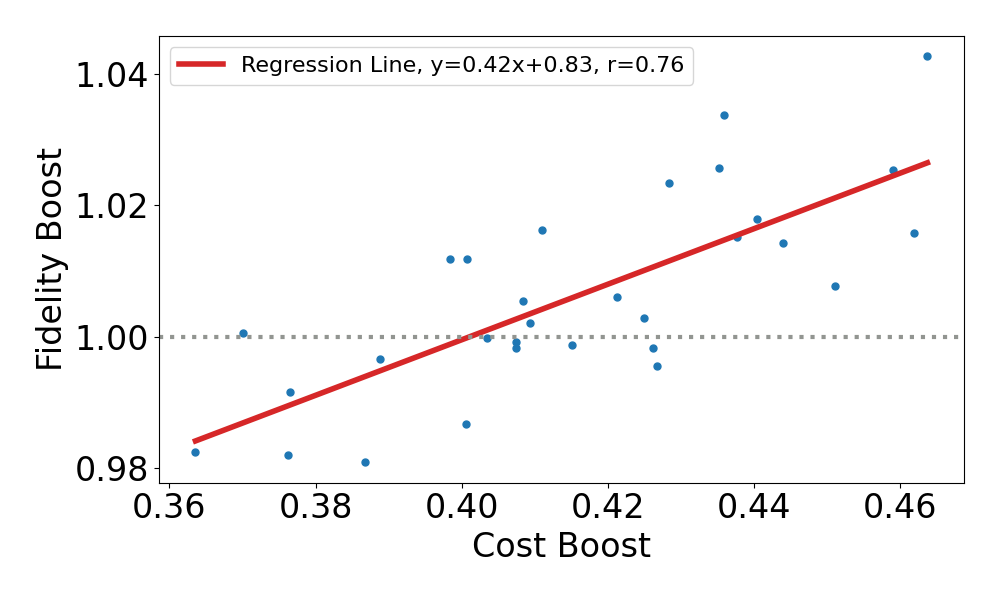}
  \caption{Correlation between the improvement of state fidelity from unmitigated to mitigated with the improvement to cost evaluation from unmitigated to mitigated for $N=6$}
  \label{fig:N6_Correlation}
\end{figure}

\begin{figure}[h]
  \centering
  \includegraphics[width=.49\textwidth]{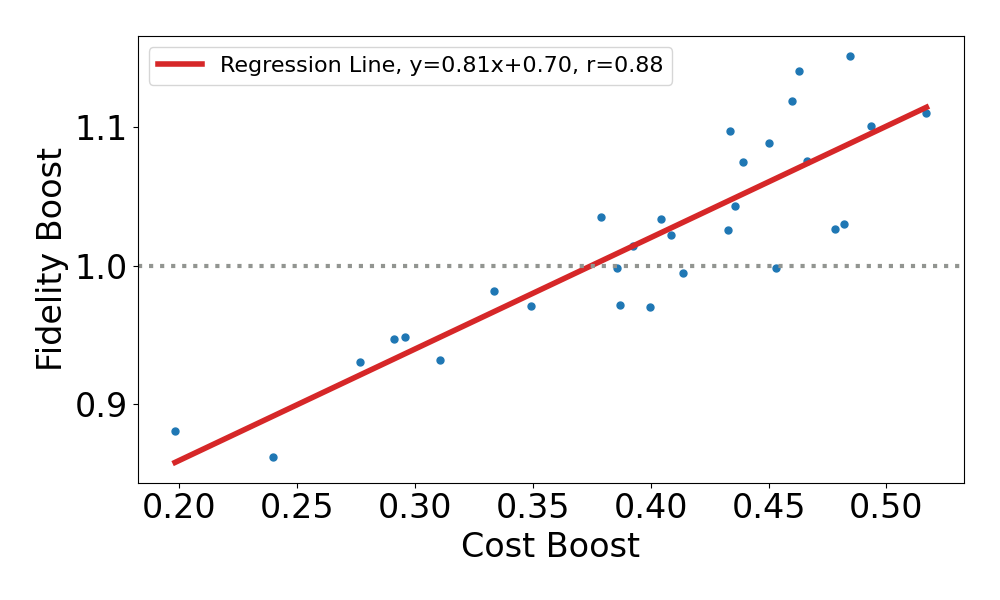}
  \caption{Correlation between the improvement of state fidelity from unmitigated to mitigated with the improvement to cost evaluation from unmitigated to mitigated for $N=8$}
  \label{fig:N8_Correlation}
\end{figure}

Also quantifying the improvement to cost evaluation as,

\begin{equation}
B_{cost} = \frac{\mathrm{Tr}[H\rho]-\mathrm{Tr}[H\rho_{em}]}{\mathrm{Tr}[H\rho]-\mathrm{Tr}[H\rho_{em}]}
\end{equation}

it can be seen that there is a correlation between the two, as shown in figures \ref{fig:N4_Correlation}, \ref{fig:N6_Correlation}, and \ref{fig:N8_Correlation}. This suggests that by using repeated runs and taking the result with the best cost improvement, one could be more confident that there is also an improvement to fidelity. However, this further increases the sampling overhead required.

\section{Adiabatic State Preparation, \\ Symmetry Verification, \\ and White Noise Calibration}

Adiabatic quantum algorithms are by now standard methods for the preparation of many-body ground states, where preparation involves the evolution of some simple initial state by a continuously-varying Hamiltonian. Indeed, the method was first popularised in the quantum computing community for use in combinatorial search problems (e.g. $n$-bit satisfiability) \cite{Farhi2000a,Farhi2000b}. Quickly adapted as a molecular state preparation method by quantum chemists \cite{Aspuru2005,Bauer2020}, the general technique has become known as adiabatic state preparation (ASP), and has also gained popularity among condensed-matter theorists \cite{Yarloo2024,Lutz2025}.

Within the high-energy physics and lattice gauge theory communities, ASP has been most fully applied to the $(1+1)-d$ Schwinger model detailed in Section~\ref{Sec:Schw}. In particular, Refs.~\cite{Chakraborty2022,Honda2022} showed how to prepare Schwinger ground states in the presence of theta (topological) terms and probe charge, and how to recover the expectation value of the chiral condensate, a useful order parameter. Even within the past year, improved ASP protocols have been developed to study string breaking and phases of the $(1+1)-d$ Schwinger model \cite{Danna2025}, including by preparation of multiple low-lying states \cite{Kaikov2025}, and targeted highly-excited states \cite{Hwang2025}.

Following \cite{Albash2018,Bauer2020,Lutz2025}, we now review the standard ASP construction and error bounds. The adiabatic theorem \cite{Albash2018} states that an initial ground state evolving under a parameterised Hamiltonian $H(\lambda)$ will remain in the ground state for all $\lambda$ so long as (i) there is no $\lambda$ where the gap $\Delta(\lambda) \equiv E_1(\lambda) - E_0(\lambda)$ vanishes, and (ii) the evolution is slow, i.e. continuous. For a digital quantum simulation, the evolution will of course never be continuous; $H(\lambda)$ must be somehow discretised over physical time. Fortunately, it is possible to strictly constrain how much error will accumulate on the ground state prepared in this way. The gap requirement is stricter, but can be overcome with care, as performed e.g. in \cite{Kaikov2025} for smooth level-crossings.

The time required for adiabatic state preparation is controlled by both the slowness of the Hamiltonian path and the minimum spectral gap $\min_\lambda|\Delta(\lambda)|$ along that path. Let $\varepsilon = \max_{s \in [0,1]} \bigl\| \tfrac{dH}{ds}(s) \bigr\| $ denote the maximal norm of the Hamiltonian derivative. The required adiabatic evolution time satisfies a bound of the form $T \gtrsim \varepsilon / g_{\min}^2$~\cite{Bauer2020}. So long as the ratio $\varepsilon/g_{\min}^2$ grows only polynomially with system size, this bound ensure that simulation by adiabatic state preparation will incur only polynomial cost. 

For the lattice Schwinger model Hamiltonian, both ingredients are favorable. Away from the critical line for $\theta = \pi$, the model remains gapped in the continuum limit, with a mass gap of order $g/\sqrt{\pi}$, implying $g_{\min} = \Theta(1)$ along the chosen interpolation path. Second, the schedule used in the adiabatic evolution only scales local hopping and staggered-field terms, so that $\|dH/ds\|$ is set by a sum over $O(N)$ local terms. We therefore roughly expect a linear growth of the adiabatic runtime $T = O(N)$ with lattice size $N$.

One convenience of the ASP technique, in comparison to variational state preparation, is that at no point will we have to evaluate the Hamiltonian observable on our prepared states (unless we are interested in energies/gaps themselves). For example, we may be interested in a standard order parameter like the chiral condensate:
\begin{equation}
    \mathcal{C} = \frac{1}{N} \left( \sum_{n=1}^{N} (-1)^n \hat{\sigma}^z_n \right),
\label{chiralcond}
\end{equation}
which is simply the (staggered) averaged expectation value of $\hat{\sigma}^z$ up to an irrelevant constant. 
Since this observable is mutually-diagonalisable with both our symmetries of note (parity, and Baryon number conservation), we are in the fortunate position that both expectation values may be extracted from a single measurement on all qubits.

We perform our error mitigation with ASP, therefore, by measuring all qubits in the computational basis, and then simply discarding those measurements which lie outside the subbasis for the desired symmetry sector. For example, for $N=4$, the the $B=0$ sector is spanned by those basis vectors with occupation number 2, i.e. $\{ \ket{0011}, \ket{0110}, \ket{1100}, \ket{1001}, \ket{0101}, \ket{1010} \}$; the parity-even sector is spanned by all these in addition to the extremal-occupation states $\{ \ket{0000}, \ket{1111} \}$.

\begin{figure}[t!]
  \centering
  \includegraphics[width=.49\textwidth]{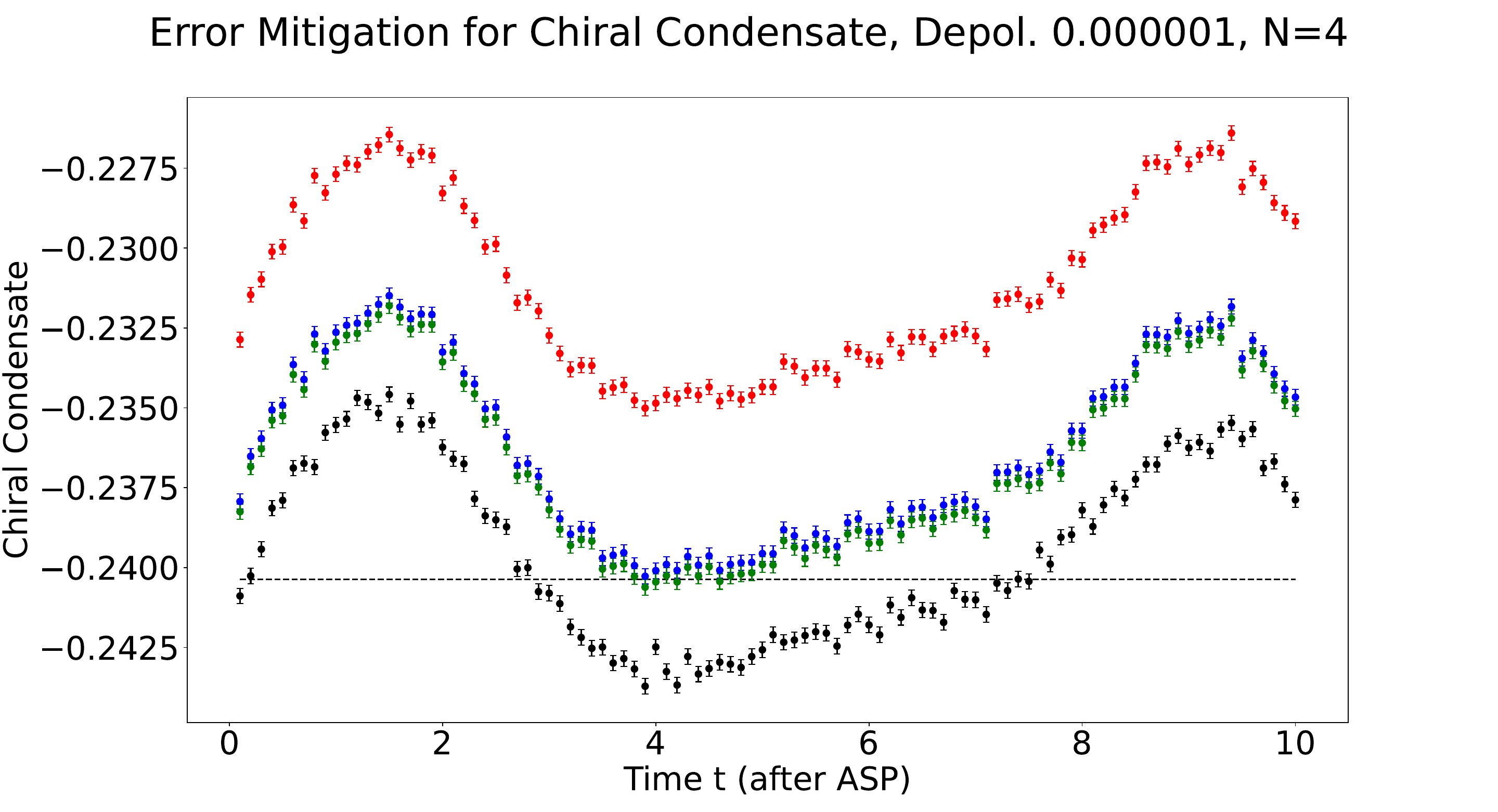}
  \caption{Adiabatic state preparation performance with and without symmetric error-mitigation, for $N=4$ sites. }
  \label{fig:adiabatic_n4}
\end{figure}

\begin{figure}[t!]
  \centering
  \includegraphics[width=.49\textwidth]{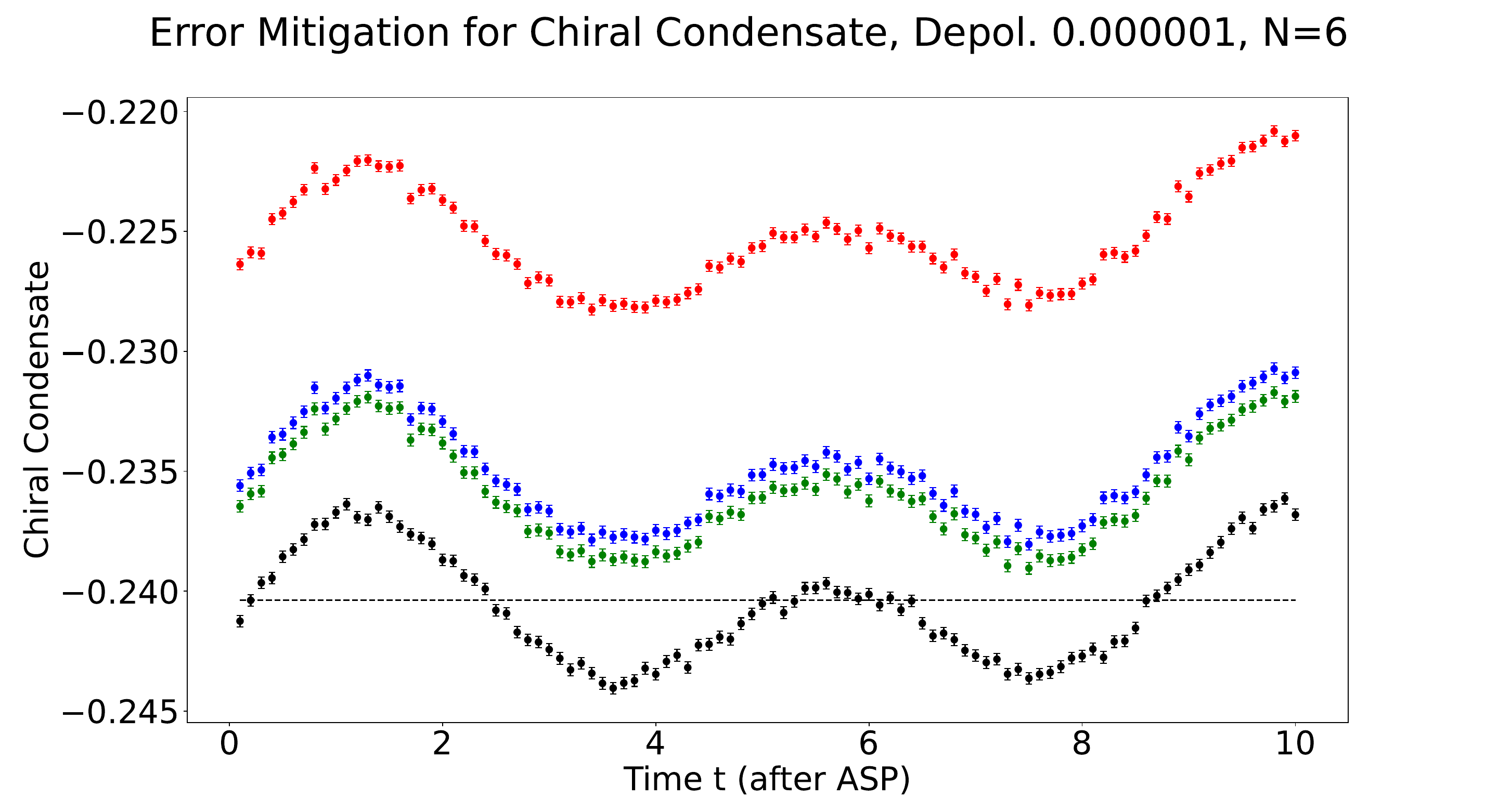}
  \caption{Adiabatic state preparation performance with and without symmetric error-mitigation, for $N=6$ sites. }
  \label{fig:adiabatic_n6}
\end{figure}

As a first demonstration, we emulate ASP for duration $T=100$ with timesteps $dt=0.1$ (1000 adiabatic steps), and then time-evolve the prepared state for $100$ steps at the same $dt$. The results are shown in Figure~\ref{fig:adiabatic_n4} for $N=4$ sites, and Figure~\ref{fig:adiabatic_n6} for $N=6$ sites. At the chosen physics parameters $\{ m=0, g=1, a=1, \theta = 0\}$, a chiral condensate expectation value of about $-0.24$ is expected, and in both cases the noiseless (black) preparation  is seen to oscillate around this value, with shot-based perturbations far smaller than the physical oscillation due to the imperfect ground-state. Depolarising hardware noise at rate $p=10^{-6}$ is then applied, and the resultant noisy (red) prepared state has a chiral condensate which is offset in the positive direction, by about $4\%$ for $N=4$ and $6\%$ for $N=6$. 

We emphasize that each datapoint for distinct evolution times $t$ is collected by a distinct emulated run of the quantum circuit, with $n=10^6$ shots. Therefore, the similarity in evolution shapes between the noisy and noiseless runs informs us that much of the hardware noise has acted globally; this will inform our global noise correction below.

Figures~\ref{fig:adiabatic_n4} and \ref{fig:adiabatic_n6} also illustrate the chiral condensate expectation values when symmetric error mitigation is applied to the noisy emulator results, for both the parity symmetry (blue) and the Baryon number symmetry (green). They have the effect of recovering about $60\%$ of the precision lost to hardware noise; the parity symmetry has only slightly less of an effect, because the extremal-occupation states $\{ \ket{0000}, \ket{1111} \}$ are less likely to be inadvertently prepared due to hardware errors.

One immediate realisation from the above results is the seemingly constant improvement ratio between unmitigated, mitigated, and noiseless chiral condensate expectation values. This ratio is essentially constant across evolution time (after ASP), and also near-constant for different lattice sizes, to an extent which we will characterise. This is a surprising feature of our analysis, for we have already seen that energy observables, at least in the case of the variational approach, do not afford such a linear and predictable improvement with symmetry-based error mitigation.

The reason for this regularity may be found in \cite{Foldager_2023}: sufficiently deep quantum circuits can, under certain generic conditions, scramble local noise into a global white noise pattern. The white noise in question may be well-approximated by the channel:
\begin{equation}
    \rho_{wn} = \eta \rho_{ideal} + (1-\eta) \mathbb{I}/d,
\end{equation}
where $\eta$ is a global error scale factor related to the channel fidelity by $F = \eta + \frac{1-\eta}{d} \simeq \eta$. The practical significance of this observation is that, in principle, the scale factor $\eta$ may be learned from a combination of quantum and classical data, and then applied in regimes where classical data is unavailable.

\begin{figure}[t!]
  \centering
  \includegraphics[width=.49\textwidth]{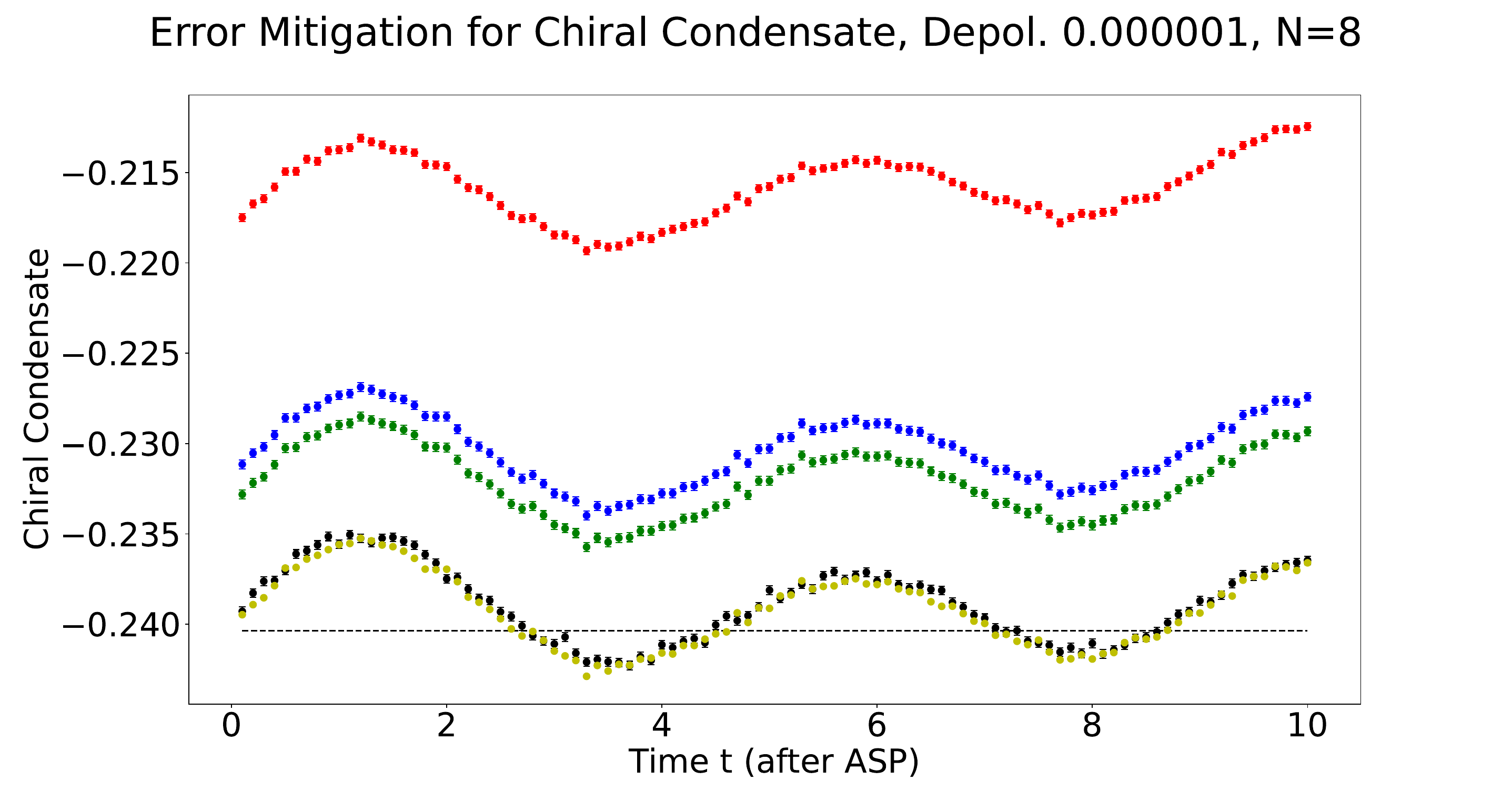}
  \caption{Adiabatic state preparation performance with and without symmetric error-mitigation, now with global noise fit (yellow), for $N=8$ sites. }
  \label{fig:adiabatic_n8_whitenoise}
\end{figure}

\begin{figure}[t!]
  \centering
  \includegraphics[width=.49\textwidth]{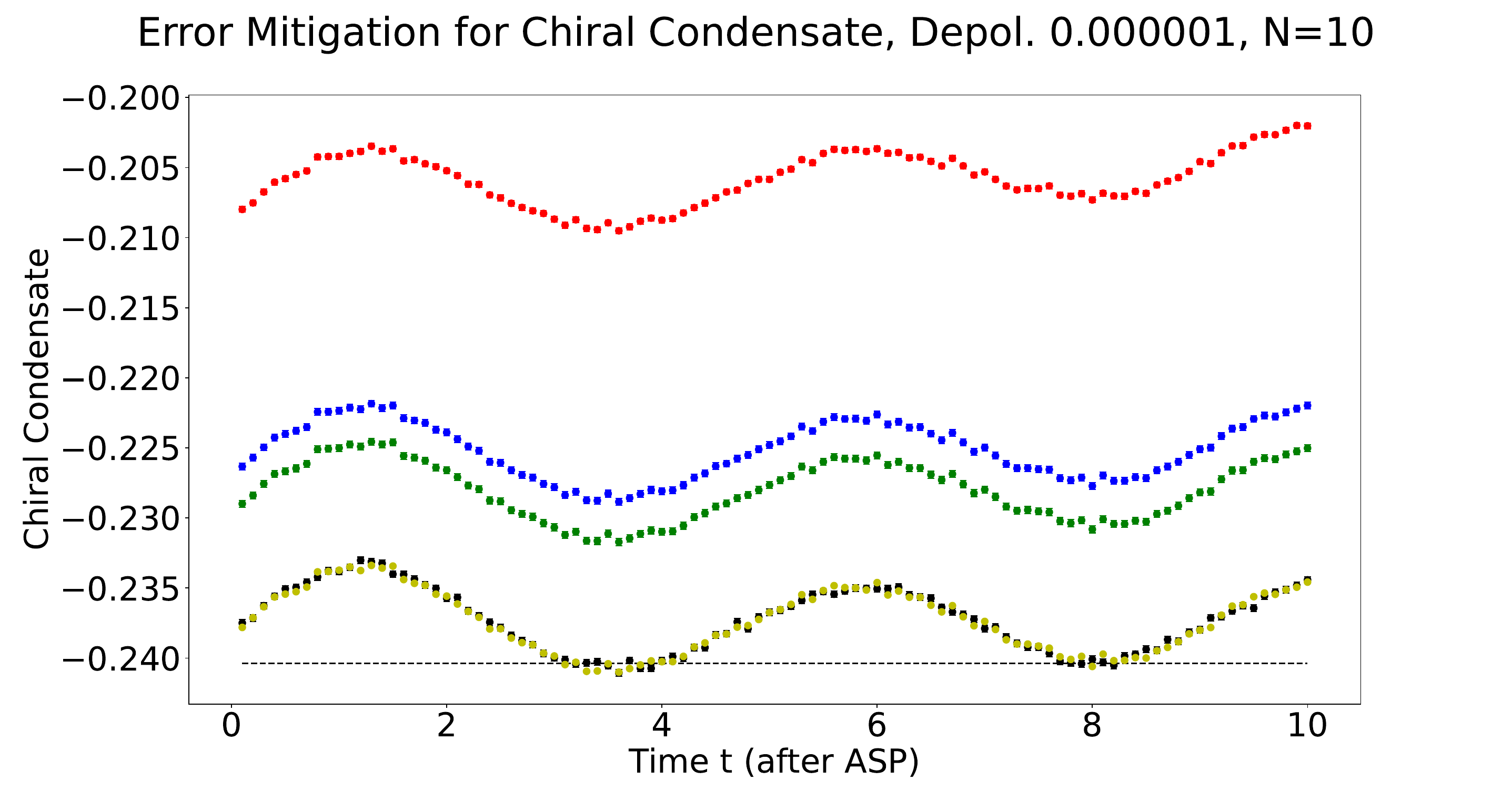}
  \caption{Adiabatic state preparation performance with and without symmetric error-mitigation, now with global noise fit (yellow), for $N=10$ sites. }
  \label{fig:adiabatic_n10_whitenoise}
\end{figure}


To define our fitting scheme, we propose simply that the improvement ratio between unmitigated, symmetry-verified, and noiseless (classical) chiral condensate expectation values is linearly related to this $\eta$. By performing a fit of the time-averaged improvement ratio for each lattice size $N$, we can then (linearly or otherwise) extrapolate to larger $N$ and attempt to reconstruct the noiseless chiral condensate values from the pair of unmitigated and symmetry-verified values which are readily available.

As a first demonstration, we use the emulated $N=4,6$ data seen already in Figures~\ref{fig:adiabatic_n4} and \ref{fig:adiabatic_n6}, to learn the improvement ratio, and then apply this ratio to the analogous $N=8,10$ data as seen in Figures~\ref{fig:adiabatic_n8_whitenoise} and \ref{fig:adiabatic_n10_whitenoise}. In addition to noiseless, noisy, and error-mitigated chiral condensate expectation values, in these plots we also include the global-noise corrected values (yellow). In this very low-hardware-noise regime, the method works flawlessly, with the improvement ratio handily recovering the noiseless values within shot-based uncertainty. For all evolution times, the two are essentially indistinguishable.

\begin{figure}[t]
  \centering
  \includegraphics[width=.49\textwidth]{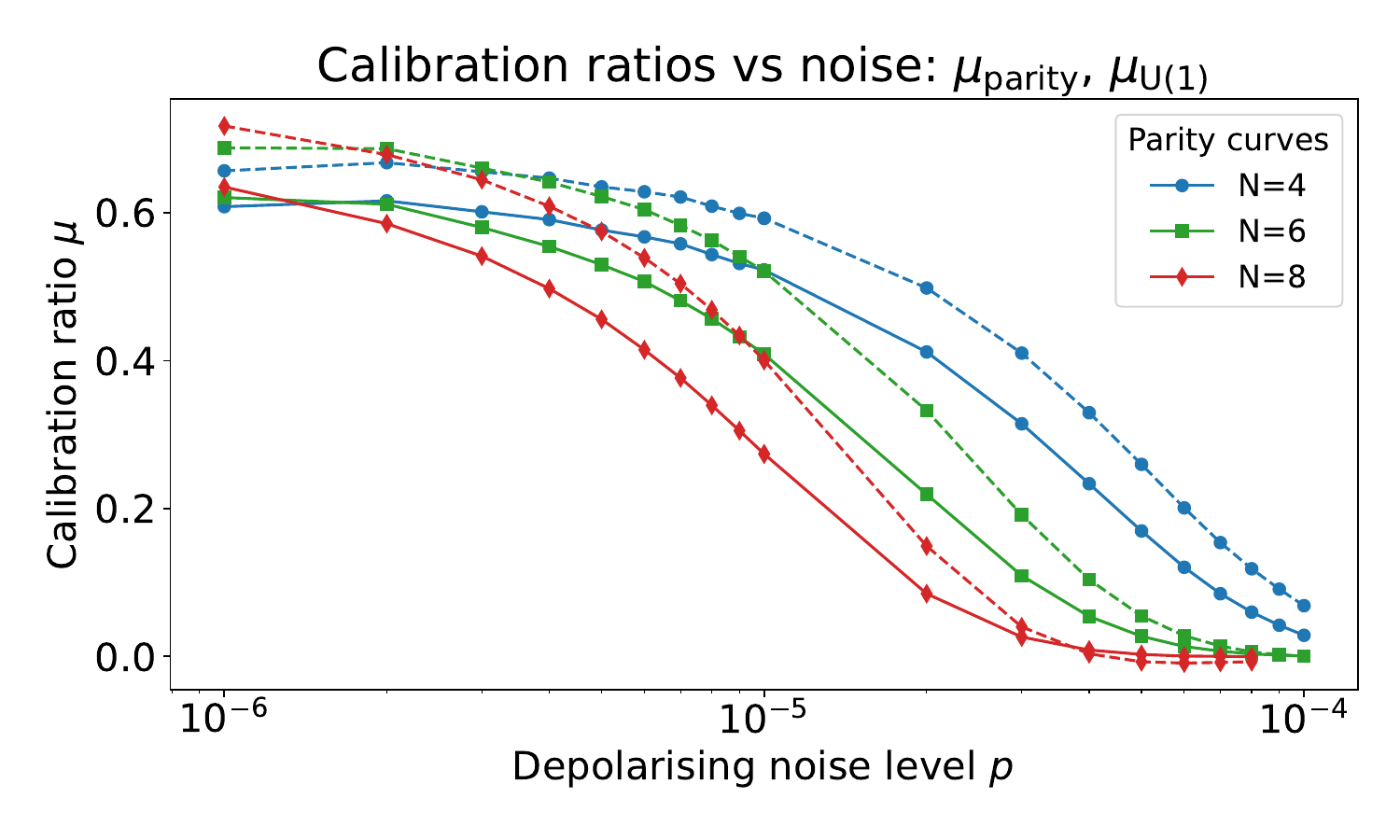}
  \caption{Global-noise calibration ratio against noise level, for $N=4,6,8$ sites. }
  \label{fig:CR_n468}
\end{figure}

It is of course important to understand how this technique fares as hardware noise approaches more realistic near-term levels. In Figure~\ref{fig:CR_n468}, we study how the improvement ratio varies between depolarizing noises $p \in [10^{-6}, 10^{-4}]$, for each of $N=4,6,8$ sites. Unsurprisingly, the trend is monotonic downward, as symmetric error mitigation achieves comparatively less for worse hardware noise, as well as for more sites. However in our global noise calibration, it is not the absolute improvement ratio itself which matters, but rather our ability to predict e.g. the $N=8$ improvement ratio given the $N=4,6$ ratios at the same noise level.

\begin{figure}[t]
  \centering
  \includegraphics[width=.49\textwidth]{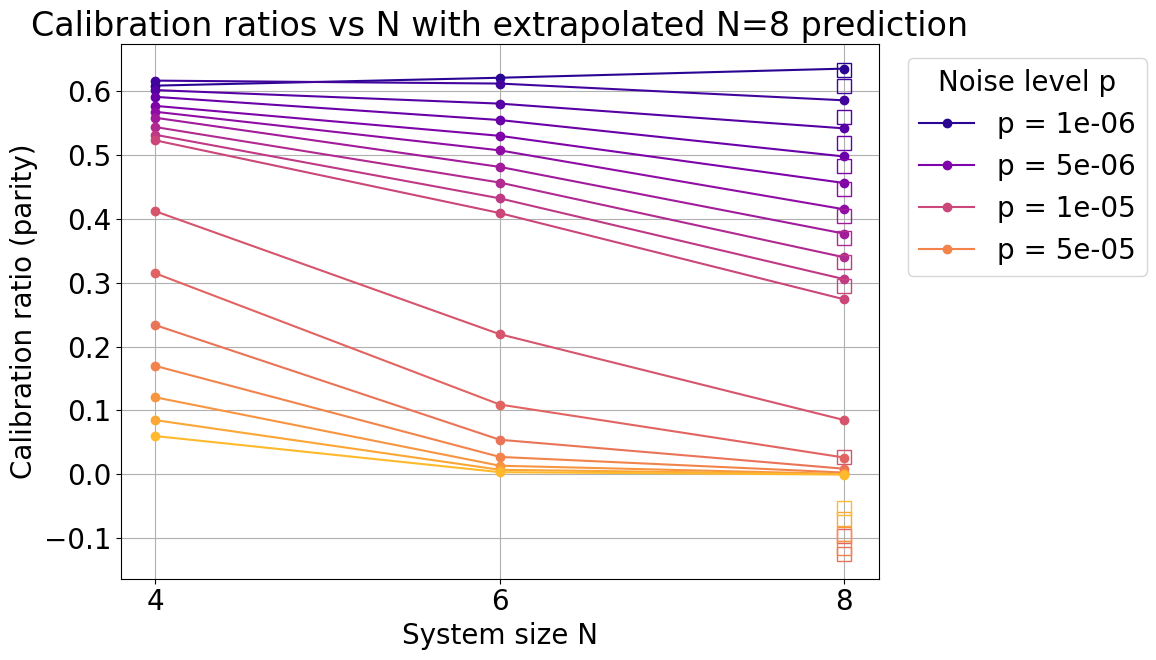}
  \caption{Calibration ratio against number of sites $N$, exhibiting failure of linear fit above $p \simeq 10^{-5}$ }
  \label{fig:crs_n468}
\end{figure}

To this end, in Figure~\ref{fig:crs_n468} we perform a simple linear extrapolation of the $N=4,6$ ratios to examine how well they recover the $N=8$ ratios, for the same range of depolarizing noise levels. The match is within $\sim 10\%$ up to $p=10^{-5}$, but degrades rapidly for higher hardware noise as the trend over $N$ gains nonlinearity. Of course, even with current resource limitations, one could fit an appropriate nonlinear model to e.g. all numbers of sites between $N \in [4,16]$, and so predict the improvement ratios for $N>16$ where emulation is inaccessible.

\begin{figure}[t]
  \centering
  \includegraphics[width=.49\textwidth]{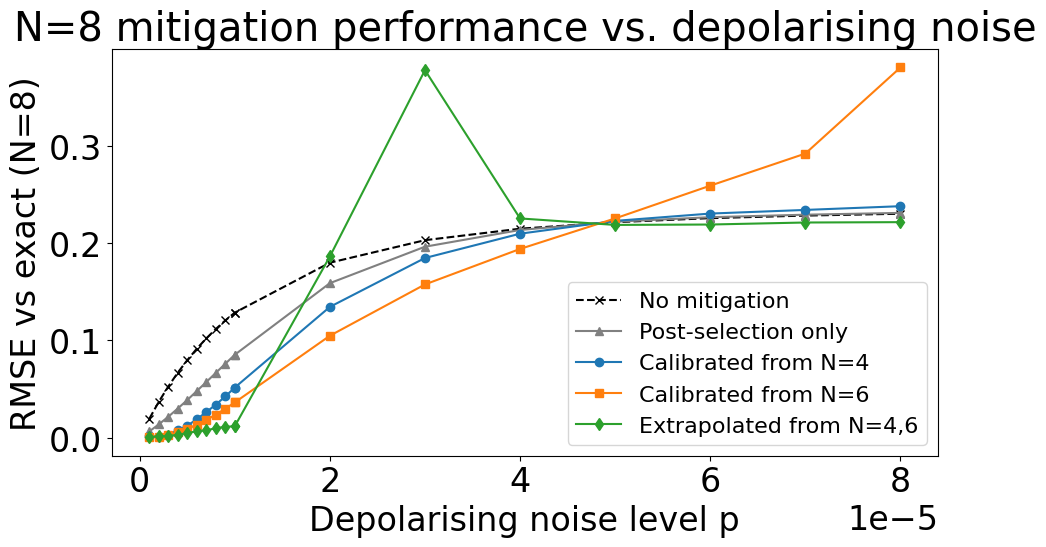}
  \caption{Chiral condensate RMSE after global-noise calibration ratio, scanned against noise level for $N=8$ sites. }
  \label{fig:CR_n8}
\end{figure}

Even with this limited linear fit, we can obtain significant error mitigation even for hardware noise up to $p=10^{-5}$, beyond which it becomes advantageous to simply take the learned $N=6$ improvement ratio as our estimate for the $N=8$ ratio, up to $p=5\cdot 10^{-5}$. Figure~\ref{fig:CR_n8} depicts how the RMS error is reduced for $N=8$ from the noise case (black), first by our symmetric error mitigation (grey), and subsequently by employing the improvement ratio. For comparison purposes we calibrate with the $N=4$ improvement ratio (blue), with the $N=6$ improvement ratio (blue), and with the linear-fit extrapolation (green). Regardless of which strategy is taken, there is a critical noise level $p^* \sim 5\cdot10^{-5}$ beyond which the global noise correction is impossible; indeed for $p > p^*$, the amount of error mitigation from symmetry postselection is essentially nil.

\begin{figure}[t]
  \centering
  \includegraphics[width=.49\textwidth]{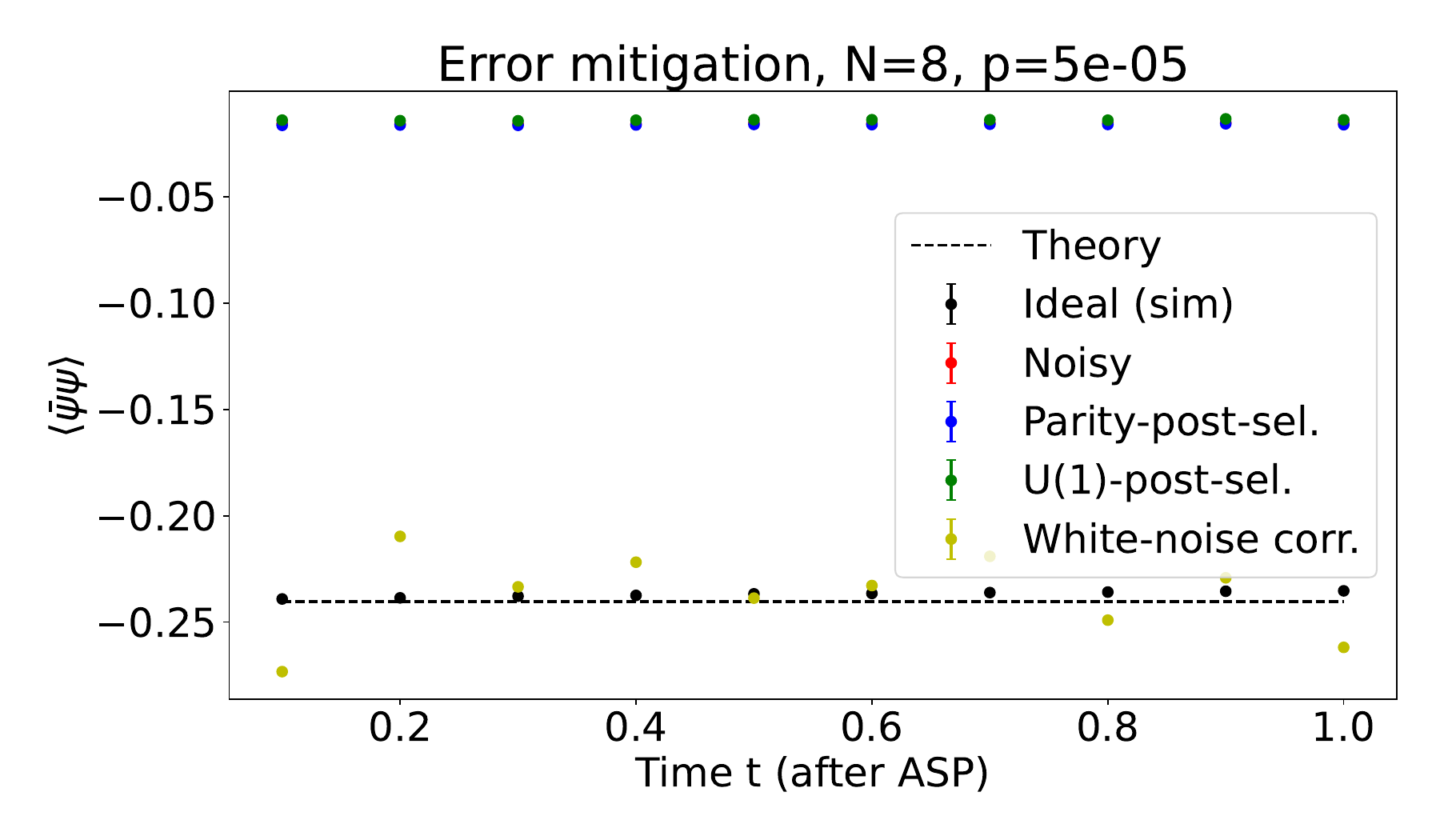}
  \caption{Time-series with global noise calibration at critical noise level $p=5 e-5$, for $N=8$ sites. }
  \label{fig:ts_n8_critnoise}
\end{figure}

As a final example of the worst-case scenario where the global noise correction has any success, we show in Figure \ref{fig:ts_n8_critnoise} the time-evolved chiral condensates for the noiseless (black) and noisy emulations (red), with symmetry error-mitigation (blue/green), and with the $N=6$ global noise improvement ratio applied (yellow). Here the noisy and symmetry-mitigated data are nearly indistinguishable to the eye; nevertheless, the improvement ratio is consistent enough that the corrected values still land much nearer to the ground truth, and are unbiased. However, a slightly larger $p$ leads to an explosion in the variance of these corrected values, such that even if they are unbiased, no individual value is trustworthy.

\section{Discussion and Conclusion}

Quantum error mitigation via symmetry verification is neither generically useful nor generically useless when applied to quantum simulations of lattice gauge theories. We have explored the practical boundaries of its beneficial use through two investigations of the $(1+1)$-dimensional Schwinger model: (1) parity- and number-sector verification during VQE optimisation; and (2) parity- and number-sector postselection for an adiabatically prepared state and its subsequent chiral condensate dynamics. In both investigations, we sought to establish whether and when symmetry information could be exploited to reduce the error in the intended lattice gauge theory calculation.

For VQE, we observed that while symmetry verification could indeed result in a better estimate of the ground-state energy, it did not generally produce better variational parameters or a higher-fidelity learned state. This is a prime example of the extent to which energetic performance can mask true state preparation performance.

Activating symmetry verification specifically produced an immediate improvement in the estimated energy for fixed variational parameters, thereby correcting the estimator evaluated at that particular point in parameter space. It did not, however, tend to initiate a new period of performance improvement, and independently optimised runs with and without this error mitigation ultimately prepared states of similar fidelity. We note, of course, that results from small problem sizes $N=4,6,8$ are not guaranteed to extrapolate; in our particular case, the choice of a symmetry-preserving ansatz which eliminates redundancies could reasonably be expected to make the variational landscape more robust than a generic ansatz. Larger or more intricate landscapes could well respond differently to in-the-loop symmetry verification, and we believe this to be worthy of further study.

On the other hand, postselection in the adiabatic calculation consistently reduced the estimation bias of the chiral condensate, eliminating as much as $\sim60\%$ of the noise-induced bias for both tested system sizes $N=4,6$. Symmetry filtering was effective at removing contributions from states that had detectably leaked out of the target symmetry sector. We found, however, that the benefits decayed rapidly as the per-gate noise probability approached an empirical crossover around $5 \times 10^{-5}$. This value should not be interpreted as a universal threshold for symmetry verification: it depends on the model, circuit family and depth, target observable, noise model, and postselection procedure used here. Since this regime is more demanding than that typically available for present-day circuits, our results illustrate how symmetry-guided error mitigation may become useful for quantum simulations of lattice gauge theories as hardware fidelity improves.

We ultimately proposed a calibration method, trained on classically accessible systems, through which the remaining chiral condensate bias could be reduced further. This method was motivated by our observation that the fraction of the bias removed by postselection was nearly time-independent and varied smoothly with problem size. We therefore estimated this relationship using the $N=4,6$ systems and transferred the resulting calibration to the held-out $N=8,10$ systems. In the lowest-noise regime tested, the calibrated estimates agreed closely with the corresponding noiseless condensates.

One lesson to be taken from our results is that an identical error mitigation strategy can have qualitatively different impacts depending on whether or not it is deployed inside a variational feedback loop. We would conjecture that this is a general phenomenon, because mitigation of an intermediate quantity is valuable only to the extent that it alters the final state of the algorithm. A lower-bias cost value at a single iteration in an optimizer is only beneficial if it changes the \textit{trajectory} of the optimizer. Further investigations would need to examine optimization trajectories directly, including changes in gradients which result from symmetry verification.

Our VQE results do not exclude all possible benefits from in-the-loop symmetry verification; indeed they point towards either intermittent mitigation, e.g. every $k$ iterations, or some adaptive mitigation policy, as strategies which might exceed the performance floor we have established.

Further studies of our residual-noise calibration step will primarily assess how the method scales with a greater number of classically-accessible system sizes. It would also be valuable to establish whether randomized compiling e.g. Pauli twirling could make the residual noise more nearly global, and in turn improve the stability of the calibration. Ultimately, the method must be tested on near-term quantum error to understand its performance in the presence of inhomogeneous noise.

There are numerous LGT observables of current physical interest which could benefit from the methods proposed here, including confinement and string-breaking, charge correlations, finite-density sectors as in LGTs with matter fermions, and real-time processes. It would be especially interesting to examine symmetry-verification performance on gauge theories which include both local and global constraints; in two or more spatial dimensions, one cannot generically fermionize, and so error mitigations which are mindful of local symmetries could enact significant benefits.

We have illustrated how symmetry verification may be advantageously deployed on quantum simulation algorithms for lattice gauge theories. For symmetry-compatible observables such as the chiral condensate, postselection can give a low-overhead bias reduction which is immediately useful, and which may also feed a global-noise calibration as demonstrated here. For near-term variational algorithms, we draw a more cautious conclusion that in-the-loop symmetry verification may improve the apparent energy without improving the prepared ground state itself. Our results nevertheless motivate the search for a ``Goldilocks'' in-the-loop error mitigation scheme, which uses symmetry verification in an informed manner to guide the trajectory of the optimizer towards a more precise final state.

\section*{Acknowledgement}
We would like to thank Prof. B\'{a}lint Koczor, Dr. Zong-Gang Mou. Research of BC at the University of Southampton
has been supported by the following research grants - STFC (Grant no. ST/X000583/1), STFC (Grant no. ST/W006251/1), and EPSRC (Grant no. EP/W032635/1). Research of GVG at the University of Southampton has been supported by STFC (Grant no. ST/X000583/1). Research of AT at Southampton has been supported by Mayflower Scholarship. ZC acknowledges support from the EPSRC Quantum Technologies Career Acceleration Fellowship (UKRI1226). We would like to acknowledge the quantum computational hardware resources provided Iridis at the University of Southampton, and by National Quantum Computing Center (NQCC) and IBM under NQCC's Quantum Computing Access Programme (QCAP). We have used QISKIT software for this work. 

\bibliographystyle{plainnat}
\bibliography{bib}

\end{document}